\documentclass[letterpaper, 10 pt, conference]{ieeeconf}  

\IEEEoverridecommandlockouts                              
\usepackage{mathtools}
\usepackage{amsmath}
\usepackage{amssymb}
\usepackage{mathrsfs}
\usepackage{graphicx}
\usepackage{subfigure}
\usepackage{color}
\usepackage{algorithm}
\usepackage{algorithmic}

\newtheorem{asmp}{\textbf{Assumption}}

\newtheorem{thm}{\textbf{Theorem}}
\newtheorem{rmk}{\textbf{Remark}}

\allowdisplaybreaks[4]

\title{\LARGE \bf
Community Detection for Gossip Dynamics with Stubborn Agents
}

\author{Yu Xing, Xingkang He, Haitao Fang, and Karl H. Johansson
\thanks{This work is supported by National Key R\&D Program of China (2016YFB0901900), National Natural Science Foundation of China (61573345), Knut \& Alice Wallenberg Foundation, and Swedish Research Council.}
\thanks{Yu Xing, Xingkang He, and Karl H. Johansson are with Division of Decision and Control Systems, School of Electrical Engineering and Computer Science, KTH Royal Institute of Technology, SE-10044 Stockholm, Sweden.
        {\tt\small yuxing2@kth.se; xingkang@kth.se; kallej@kth.se}}%
\thanks{Haitao Fang is with Key Laboratory of Systems and Control, Academy of Mathematics and Systems Science, Chinese Academy of Sciences, Beijing 100190, and School of Mathematical Sciences, University of Chinese Academy of Sciences, Beijing 100049, P. R. China. 
        {\tt\small htfang@iss.ac.cn}}%
}

\begin{document}

\maketitle

\begin{abstract}
We consider a community detection problem for gossip dynamics with stubborn agents in this paper. It is assumed that the communication probability matrix for agent pairs has a block structure. More specifically, we assume that the network can be divided into two communities, and the communication probability of two agents depends on whether they are in the same community. Stability of the model is investigated, and expectation of stationary distribution is characterized, indicating under the block assumption, the stationary behaviors of agents in the same community are similar. It is also shown that agents in different communities display distinct behaviors if and only if state averages of stubborn agents in different communities are not identical. A community detection algorithm is then proposed to recover community structure and to estimate communication probability parameters. It is verified that the community detection part converges in finite time, and the parameter estimation part converges almost surely. Simulations are given to illustrate algorithm performance.
\end{abstract}

\section{INTRODUCTION}

Community detection is a fascinating topic of network science, which has attracted attention of researchers from multiple disciplines for decades \cite{fortunato2016community, schaub2017many}. Its goal is to identify similar nodes in a network based on their connections and behaviors, for example, finding protein groups having the same function in protein regulatory networks, and discovering websites with related topics in World Wide Web \cite{fortunato2010community}. As a consequence of its wide existence and applications, a great number of approaches have been well-studied, such as spectral clustering \cite{von2007tutorial}, modularity optimization \cite{newman2004finding}, and statistical inference methods for generative models \cite{abbe2017community}.

Recently, there is a growing interest in community detection for dynamical systems in control society and other domains \cite{prokhorenkova2019learning, peixoto2019network, wai2019blind, schaub2019blind, roddenberry2020exact, ramezani2018community}. This stands in stark contrast to classic community detection problems where networks can be directly observed in general. The fact that only states of nodes can be obtained, rather than edge sets, complicates the issue significantly. In \cite{prokhorenkova2019learning, peixoto2019network,ramezani2018community}, statistical inference methods were used to solve community detection for diffusion processes, while spectral methods were introduced in \cite{wai2019blind, schaub2019blind, roddenberry2020exact}. 

A natural way to solve the community detection problem is to follow a two-step procedure which first recovers underlying networks in some sense and then clusters nodes based on the estimation. For example, \cite{prokhorenkova2019learning} constructed weighted graphs from diffusion dynamics in several ways and then applied Louvain algorithm to find clusters, while authors in \cite{ramezani2018community} used behavior similarity between nodes to represent the underlying graph and then addressed the problem via a maximum likelihood approach. However, it may be difficult to estimate the underlying networks with desired accuracy, which could result in degraded performance of the community detection step, if there is insufficient excitation for the system. Hence a key issue is how to obtain community structure, which is the correspondence between nodes and communities, indicating which community a node belongs to, by directly using state observations. 

One framework to deal with this problem, referred to as blind community detection, is introduced in \cite{wai2019blind, schaub2019blind, roddenberry2020exact}, where state observations are used to compute sample covariance matrices providing information of community structure. But in these researches, several realizations of the considered systems are necessary, as well as sufficient excitation for initial values of different instances. These may not be satisfied for realistic scenarios such as discussions and innovation diffusion, where a process could only happen once, or the initial values of different instances may be highly correlated. So it is still a question whether one can address detection problems online, based on single trajectory. 

In this paper, we consider community detection for gossip models with stubborn agents. Gossip models have been extensively studied in control society for their application in consensus algorithms \cite{fagnani2008randomized} and modeling opinion formation processes \cite{acemouglu2013opinion}. In \cite{acemouglu2013opinion}, the authors showed that the existence of stubborn agents, which never change their states, leads to persistent fluctuation of the dynamics. We would like to investigate whether one can identify the community structure of this type of processes by only observing states of agents. Since gossip models are widely used in modeling social opinion formation, learning community structure based on state observations can reveal key information of social structure of the underlying group.

It is assumed in this paper that the communication probability matrix for agent pairs has a block structure. More specifically, agents can be divided into two communities, and the communication probability of two agents depends on whether they are in the same community. This simplifies the model, but difficulty still remains since we do not know the community structure. 

There are two key differences between our paper and previous studies. First, the paper focuses on community detection for gossip dynamics by using properties of states directly, rather than utilizing intermediate estimations of underlying networks as in \cite{prokhorenkova2019learning, ramezani2018community}. Additionally, we consider an online community detection problem, recovering the community structure gradually as the process goes on. As a result, there is no need for running multiple experiments. For example, it was necessary in \cite{wai2019blind, schaub2019blind, roddenberry2020exact} to observe a state vector at the same time step for the system starting from different initial conditions.  

Our contributions are as follows:

1. An online community detection problem is considered for gossip dynamics with stubborn agents. After assuming agents can be divided into two communities and the communication probability matrix has a block structure, we propose a recursive algorithm to recover the community structure and estimate the communication probability, based on single trajectory. It is verified that the community detection part of the algorithm converges in finite time, and the parameter estimation part converges almost surely.

2. Stability of the model is studied, and expectation of stationary distribution is characterized. The latter result indicates that under the block assumption, the stationary behaviors of agents in the same community are similar. It is also shown that agents in different communities display distinct behaviors if and only if state averages of stubborn agents in different communities are not identical. 
These are key observations and play a crucial role in proving consistency of the proposed algorithm. 

\textbf{Notation and definition.} Denote $n$-dimensional Euclidean space by $\mathbb{R}^n$, the set of $n\times m$ real matrices by $\mathbb{R}^{n\times m}$, and the set of nonnegative integers by $\mathbb{N}$. Let $\mathbf{1}_n$ be the all-one vector with dimension $n$, $\textbf{e}_i$ be the unit vector with $i$-entry being one and all other entries being zero, $I_n$ be the $n\times n$ identity matrix, and $\mathbf{0}_{n,m}$ be the $n\times m$ all-zero matrix. The subscripts of the above notations may be omitted if there is no confusion. Define $\mathbf{1}_{n_1,n_2} := \mathbf{1}_{n_1} \mathbf{1}_{n_2}^T$. 

For a matrix $A = [a_{ij}]_{1\le i,j \le n} \in \mathbb{R}^{n\times n}$, denote its $(i,j)$-th entry by $a_{ij}$. The matrix $A$ is said to be row stochastic if $a_{ij}\ge 0$ and $A\mathbf{1} = \mathbf{1}$, and to be substochastic if $a_{ij}\ge 0$ and the row sums of $A$ are not larger than one. Denote the spectral radius of matrix $A$ by $\rho(A)$, and the expectation of a random vector $X$ by $\mathbb{E}\{X\}$. The cardinality of a set $\Omega$ is denoted by $|\Omega|$. The function $\mathbb{I}_{[\textup{inequality}]}$ is the indicator function equal to one if the inequality holds, and equal to zero otherwise. We call an event happens almost surely (a.s.) if it happens with probability one. 

The rest of the paper is organized in the following way. In Section \ref{sec_problem} the community detection problem is formulated. Analysis of the model is given in Section \ref{sec_mainresults}, and then the community detection algorithm is proposed. Section \ref{sec_simulation} presents several numerical simulations, and Section \ref{sec_conclusion} concludes the paper. Proofs are omitted due to page limitation.

\section{PROBLEM FORMULATION}\label{sec_problem}

Consider an undirected graph $\mathcal{G} = (\mathcal{V}, \mathcal{E})$ with $|\mathcal{V}| = n$, and a nonnegative and symmetric matrix $W=[w_{ij}]_{1\le i,j\le n} \in \mathbb{R}^{n\times n}$ with $\mathbf{1}^T_n W \mathbf{1}_n = 1$ such that $w_{ij}\not= 0 \Leftrightarrow \{i,j\} \in \mathcal{E}$. Moreover, $\mathcal{V}$ consists of two types of agents, i.e., regular and stubborn ones, denoted by $\mathcal{V}_r$ and $\mathcal{V}_s$. Hence $\mathcal{V} = \mathcal{V}_r \cup \mathcal{V}_s$ and $\mathcal{V}_r \cap \mathcal{V}_s = \emptyset$. Each agent $i$ possesses a state $x_i(t) \in \mathbb{R}$, and the overall state vector at time $t$ is denoted by $x(t)$, obtained by stacking the states of all agents. 

The gossip process with stubborn agents and fixed initial state $x(0)$ evolves as below. At every time step $t \in \mathbb{N}$, edge $\{i, j\}$ is activated with probability $2w_{ij}$ independently of previous updates, and agents update their states according to the following rule, 
\begin{align}\label{eq_update_rule1}
    x_k(t+1) &= \begin{cases}
            \frac12 (x_i(t) + x_j(t)), & \text{ if } k \in \mathcal{V}_r \cap \{i,j\},\\
            x_k(t), & \text{ otherwise},
            \end{cases}
\end{align}
where the averaging weight is set to be $1/2$ in this model, but general weights can be considered. 

The model has been widely studied, e.g. in \cite{acemouglu2013opinion}, 
and the process degenerates to the symmetric gossip model \cite{fagnani2008randomized}, when there are no stubborn agents. 

By defining 
\begin{align}\label{eq_Rij}
    R^{ij} = \begin{cases}
    I - \frac12 (\textbf{e}_i - \textbf{e}_j)(\textbf{e}_i - \textbf{e}_j)^T, & \text{ if } i, j \in \mathcal{V}_r,\\
    I - \frac12 \textbf{e}_i(\textbf{e}_i - \textbf{e}_j)^T, & \text{ if } i \in \mathcal{V}_r,  j \in \mathcal{V}_s,\\
    I - \frac12 \textbf{e}_j(\textbf{e}_j - \textbf{e}_i)^T, & \text{ if } i \in \mathcal{V}_s,  j \in \mathcal{V}_r,\\
    I, & \text{ if } i, j \in \mathcal{V}_s,
    \end{cases}
\end{align}
and a sequence of i.i.d. $n$-dimensional random matrices $\{R(t), t\in \mathbb{N}\}$ such that
\begin{align}
    \mathbb{P}\{R(t) = R^{ij}\} = 2 w_{ij},
\end{align}
the above update rule can be written in a compact form,
\begin{align}\label{eq_update_compact}
    x(t+1) = R(t) x(t).
\end{align}
Since stubborn agents never change their states in the process from \eqref{eq_update_rule1}, we can rewrite \eqref{eq_update_compact} as follows,
\begin{align}\label{eq_update_compact_regular}
    x^r(t+1) = A(t) x^r(t) + B(t) x^s(t),
\end{align}
where $x^r(t)$ and $x^s(t)$ are the state vectors obtained by stacking the states of regular and stubborn agents respectively, and $x^s(t) \equiv x^s(0)$. In \eqref{eq_update_compact_regular}, $(A(t) ~ B(t))$ is the matrix consisting of rows in $R(t)$ which correspond to regular agents.

\begin{figure*}[!ht]
\begin{align}\label{eq_barR}
    \bar{R} &= 
    \begin{bmatrix}
    (1 - w_s n_1 - w_d n_2) I_{n_{r1}} + w_s \mathbf{1}_{n_{r1},n_{r1}} & w_s \mathbf{1}_{n_{r1},n_{s1}} & w_d \mathbf{1}_{n_{r1},n_{r2}} & w_d \mathbf{1}_{n_{r1},n_{s2}}\\
    \mathbf{0} & I_{n_{s1}} & \mathbf{0} & \mathbf{0}\\
    w_d \mathbf{1}_{n_{r2},n_{r1}} & w_d \mathbf{1}_{n_{r2},n_{s1}} & (1 - w_s n_2 - w_d n_1) I_{n_{r2}} + w_s \mathbf{1}_{n_{r2},n_{r2}} & w_s \mathbf{1}_{n_{r2},n_{s2}}\\
    \mathbf{0} & \mathbf{0} & \mathbf{0} & I_{n_{s2}}
    \end{bmatrix}
\end{align}
\begin{align}\label{eq_barA}
    \bar{A} = 
    \begin{bmatrix}
    (1 - w_s n_1 - w_d n_2) I_{n_{r1}} + w_s \mathbf{1}_{n_{r1},n_{r1}} & w_d \mathbf{1}_{n_{r1},n_{r2}}\\
    w_d \mathbf{1}_{n_{r2},n_{r1}} & (1 - w_s n_2 - w_d n_1) I_{n_{r2}} + w_s \mathbf{1}_{n_{r2},n_{r2}}
    \end{bmatrix}
\end{align}
\begin{align}\label{eq_barB}
    \bar{B} = 
    \begin{bmatrix}
    w_s \mathbf{1}_{n_{r1},n_{s1}} & w_d \mathbf{1}_{n_{r1},n_{s2}}\\
    w_d \mathbf{1}_{n_{r2},n_{s1}} & w_s \mathbf{1}_{n_{r2},n_{s2}}
    \end{bmatrix}
\end{align}
\end{figure*}

From the perspective of community detection, we would like to divide the agents into different groups. This could be done if we estimate the probability matrix $W$ or $\mathbb{E}\{R(t)\} = \sum_{1 \le i, j \le n} w_{ij}R^{ij}$, a function of $W$, and then extract community information from them via traditional methods. However, these matrices may not be recovered for the considered gossip model, since there is no extra excitation in the system. A way to look at this point is to consider \eqref{eq_update_compact_regular} as a linear system with random noise,
\begin{align*}
    x^r(t+1) = \bar{A} x^r(t) + \bar{B} x^s(t) + w(t),
\end{align*}
where $\bar{A} = \mathbb{E}\{A(t)\}$, $\bar{B} = \mathbb{E}\{B(t)\}$, and $w(t) = (A(t) - \bar{A}) x^r(t) + (B(t) - \bar{B}) x^s(t)$. The data matrix $\sum_{k=0}^t\limits \left( \begin{bmatrix} x^r(k) \\ x^s(k) \end{bmatrix} [x^r(k)^T ~ x^s(k)^T] \right)$ of the least-square estimator for $(\bar{A}~ \bar{B})$ can never be invertible, if the number of stubborn agents is more than one. To see this, note that $x^s(k) = x^s(0)$ is a constant vector, so the lower part of the data matrix is just $x^s(0) (\sum_{k=0}^t [x^r(k)^T ~ x^s(k)^T])$, having only rank one. Hence, the data matrix cannot be full-rank.

In light of this intrinsic difficulty to obtain topological information, in this paper we study a simplified version of system \eqref{eq_update_compact}, to see whether community structure of the system can be obtained directly from observing system states. Here we assume that the probability matrix $W$ has a block structure, inspired by stochastic block models for community detection \cite{abbe2017community}. 
More specifically, we assume that $\mathcal{V}$ naturally has two communities $\mathcal{V}_1 = \{1, \dots, n_1\}$ and $\mathcal{V}_2 = \{n_1 + 1, \dots, n_1 + n_2\}$ with both possibly having stubborn agents, i.e., $\mathcal{V}_i = \mathcal{V}_{ri} \cup \mathcal{V}_{si}$ for $i=1,2$, and $n_1 + n_2 = n$. Here, $\mathcal{V}_{ri}$ is the set of regular agents in community $i$, and $\mathcal{V}_{si}$ the set of stubborn agents in community $i$. The numbers of agents and communities are considered to be prior information, but the cardinality of each community and the community structure are unknown and to be estimated, where a community structure is a correspondence between agents and communities, i.e., 
$\mathcal{C}(i) = k$ for $i \in \mathcal{V}_{k}$, $k=1,2$. Note that the community label is unique up to a permutation, by which we mean a redistribution of these labels to the communities. For example, we can call $\mathcal{V}_1$ as community “2", but $\mathcal{V}_2$ as community “1". In other words, the order of community labels does not matter.

Furthermore, $W$ is assumed to have the following block structure
\begin{align}\label{eq_block_W}
    \begin{bmatrix}
    w_s \mathbf{1}_{n_1,n_1} & w_d \mathbf{1}_{n_1,n_2}\\
    w_d \mathbf{1}_{n_2,n_1} & w_s \mathbf{1}_{n_2,n_2}
    \end{bmatrix},
\end{align}
so $2w_s$ (res. $2w_d$) is the probability of selecting a pair of nodes in the same community (res. different communities). This block assumption is similar to that of the symmetric stochastic block model with two communities \cite{abbe2017community}.

The problem considered in this paper is as follows. 

\textbf{Problem. } Given one trajectory of gossip dynamics \eqref{eq_update_compact}, infer the community structure of all agents, $\{\mathcal{C}(i)$, $i\in\mathcal{V}\}$, and parameters $w_s$ and $w_d$.

It is shown that in the above case the community detection problem can be solved (hence system \eqref{eq_update_compact} can be identified), if there is some prior information for stubborn agents.

\section{MAIN RESULTS}\label{sec_mainresults}

In this section, properties of system \eqref{eq_update_compact} are studied and then used to develop algorithms for community detection and parameter estimation.

\subsection{Model Analysis}
Gossip models like \eqref{eq_update_compact} have been widely studied in literature \cite{fagnani2008randomized, acemouglu2013opinion}, but we present several properties of \eqref{eq_update_compact} for completeness and further investigation. Before that, the block structure of $\bar{R} := \mathbb{E}\{R(t)\}$, $\bar{A} := \mathbb{E}\{A(t)\}$, and $\bar{B} := \mathbb{E}\{B(t)\}$ is shown in Theorem \ref{thm_barRAB}, indicating that the block structure assumption for $W$ results in similar update rules for agents in the same community. Sort regular and stubborn agents in each community in the following way for convenience, $\mathcal{V}_{r1} = \{1, \dots, n_{r1}\}$, $\mathcal{V}_{s1} = \{n_{r1} + 1, \dots, n_{1}\}$, $\mathcal{V}_{r2} = \{n_1+1, \dots, n_1 + n_{r2}\}$, and $\mathcal{V}_{s2} = \{n_1 + n_{r2} + 1, \dots, n\}$, and denote $n_{ri} := |\mathcal{V}_{ri}|$ and $n_{si} := |\mathcal{V}_{si}|$, $i=1,2$. The assumptions discussed in Section \ref{sec_problem} are summarized as follows. Note that from the update rule \eqref{eq_update_rule1}, $(A(t), B(t))$ is independent of $x(t)$ for every $t \in \mathbb{N}$.

\begin{asmp}\label{asmp_community}~\\
(i.1) $\mathcal{V}$ consists of two communities, $\mathcal{V}_1 = \{1, \dots, n_1\}$ and $\mathcal{V}_2 = \{n_1 + 1, \dots, n_1 + n_2\}$ with $n_1 + n_2 = n$. \\
(i.2) Both communities have regular agents, and there exists at least one stubborn agent in the network, i.e., $0 < n_{r1} \le n_1$, $0 < n_{r2} \le n_2$, and $n_{r1} + n_{r2} < n$.\\
(ii) The matrix $W$ has a block structure \eqref{eq_block_W} with $w_s, w_d > 0$, $w_s \not = w_d$, and $\mathbf{1}^T W \mathbf{1} = 1$.\\
(iii) The initial vector of stubborn agents $x^s(0) := \mathbf{x}^s = [(\mathbf{x}^{s1})^T ~ (\mathbf{x}^{s2})^T]^T$ is fixed.
\end{asmp}

\begin{rmk}
In Assumption \ref{asmp_community} (i.1), the order of agents is sorted for convenience, but we do not know which group they belong to before community detection. If $w_s = w_d$ holds in Assumption \ref{asmp_community} (ii), then there is no block structure at all, so it is necessary to assume they are not equal. From update rule \eqref{eq_update_compact_regular}, 
one may recover $A(t)$ and $B(t)$ by finding agents changing their states at each time. But we do not investigate this in detail, because our focus here is to recover the community structure by directly using states of agents. 
Finally, the data matrix of the least-square estimator discussed in Section \ref{sec_problem} remains non-invertible, but the below results show that the system can be identified under the above assumption, and a further one that there is some prior information for stubborn agents (Assumption \ref{asmp_stubborn_agent_structure}).
\end{rmk}

The block structures of $\bar{R} = [\bar{r}_{ij}]_{1\le i,j\le n}$, $\bar{A} = [\bar{a}_{ij}]_{1\le i,j\le n}$, and $\bar{B} = [\bar{b}_{ij}]_{1\le i,j\le n}$ are presented in the next theorem under Assumption \ref{asmp_community}.

\begin{thm}\label{thm_barRAB}
Suppose Assumption \ref{asmp_community} holds. Then $\bar{R}$, $\bar{A}$, and $\bar{B}$ have block structures presented in \eqref{eq_barR}, \eqref{eq_barA}, and \eqref{eq_barB}, respectively, where $n_{si} = n_i - n_{ri}$, $i=1,2$.
\end{thm}




Now we give the stability result of gossip process \eqref{eq_update_compact_regular}, and moreover, we show that the expectations of stationary states for regular agents in the same community are indeed the same. This provides us with possibility to recover the community structure of underlying networks.

\begin{thm}(Stability)\label{thm_stability}
Under Assumption \ref{asmp_community}, the following results hold for \eqref{eq_update_compact_regular}.\\
(i) $x^r(t)$ converges in distribution to a unique invariant distribution.\\
(ii) For any fixed initial vector $x(0)$,
\begin{align}\label{eq_expectation_limit}
    \lim_{t \to \infty} \mathbb{E}\{x^r(t)\} = (I-\bar{A})^{-1}\bar{B}x^s(0) := \mathbf{x}^r,
\end{align}
and
\begin{equation} \label{eq_ergodicity}
\lim_{t \to \infty} \frac1t \sum_{i = 0}^{t - 1} x^r(i) = \mathbf{x}^r \quad \text{a.s.}
\end{equation}
\end{thm}


In Theorem \ref{thm_stability}, $(I-\bar{A})^{-1}$ also has a block structure, as shown below. This, combined with the above theorem, indicates that the behaviors of regular agents in the same community are similar in an average sense.

\begin{thm}\label{thm_expectation_structure}
Under Assumption \ref{asmp_community}, $(I - \bar{A})^{-1}$ has the following form
\begin{align}\label{eq_inverse_I-barA}
    \begin{bmatrix}
    \frac1{a_1}(I_{n_{r1}} - \tilde{w}_{s1} \mathbf{1}_{n_{r1},n_{r1}}) & \tilde{w}_d \mathbf{1}_{n_{r1},n_{r2}} \\
    \tilde{w}_d \mathbf{1}_{n_{r2},n_{r1}} & \frac1{a_2}(I_{n_{r2}} - \tilde{w}_{s2} \mathbf{1}_{n_{r2},n_{r2}})
    \end{bmatrix},
\end{align}
where $a_i = w_s n_i + w_d n_{3-i}$, $i=1,2$, and $\tilde{w}_{s1}$, $\tilde{w}_{s2}$, and $\tilde{w}_d$ are constants depending on $w_s$, $w_d$, $n_{r1}$, and $n_{r2}$.
As a result, 
\begin{align}\nonumber
    \mathbf{x}^r &= \frac{1}{\delta}[(\gamma_{11} \mathbf{1}_{n_{s1}}^T \mathbf{x}^{s1} + \gamma_{12} \mathbf{1}_{n_{s2}}^T \mathbf{x}^{s2}) \mathbf{1}_{n_{r1}}^T, \\\nonumber
    &\quad (\gamma_{21} \mathbf{1}_{n_{s2}}^T \mathbf{x}^{s2} + \gamma_{22} \mathbf{1}_{n_{s1}}^T \mathbf{x}^{s1}) \mathbf{1}_{n_{r2}}^T]^T\\\label{eq_chi1_chi2}
    &:=  [\chi_1 \mathbf{1}_{n_{r1}}^T, ~ \chi_2 \mathbf{1}_{n_{r2}}^T]^T,
\end{align}
where
\begin{align*}
    \gamma_{11} &= w_sw_dn_1 + w_d^2n_{r2} + w_s^2n_{s2},\\
    \gamma_{12} &= w_d(w_dn_1+w_sn_2),\\
    \gamma_{21} &= w_sw_dn_2 + w_d^2n_{r1} + w_s^2n_{s1},\\
    \gamma_{22} &= w_d(w_dn_2+w_sn_1),\\
    \delta &= w_s^2n_{s1}n_{s2} + w_sw_d(n_1n_{s1}+n_2n_{s2}) \\
    &\quad + w_d^2(n_1n_2-n_{r1}n_{r2}),
\end{align*}
and $\mathbf{1}_{n_{si}}^T \mathbf{x}^{si}$ is defined to be zero if $n_{si} = 0$, $i=1,2$.
\end{thm}


The above theorem means that the expectation of stationary state of each regular agent is a weighted average of stubborn agents' states (Note that $\delta = \gamma_{11} n_{s1} + \gamma_{12} n_{s2} = \gamma_{21} n_{s2} + \gamma_{22} n_{s1}$). Moreover, the expectations for regular agents in the same community are identical. This fact makes us able to split regular agents into different groups, by observing their stationary behaviors, and leads to the design of community detection algorithm in the next subsection.

\subsection{Community Detection Algorithm}
In this subsection, we design a community detection algorithm for the problem considered in the paper. Moreover, we find conditions such that the algorithm is able to recover the community structure in finite time, and to be consistent for estimating parameters $w_s$ and $w_d$.

To partition regular agents according to their states, we introduce the following condition to ensure that $\chi_1$ and $\chi_2$ in Theorem \ref{thm_expectation_structure} are not equal. Otherwise, the regular agents exhibit similar behaviors in average, making the distinguishing task impossible.

\begin{asmp}\label{asmp_stubborn_agents}
Both communities have stubborn agents, i.e., $n_{s1}, n_{s2} > 0$, and the initial vector of stubborn agents, $\mathbf{x}^s = [(\mathbf{x}^{s1})^T ~ (\mathbf{x}^{s2})^T]^T$ satisfies 
\begin{align*}
\frac{1}{n_{s1}}\mathbf{1}_{n_{s1}}^T \mathbf{x}^{s1} \not = \frac{1}{n_{s2}}\mathbf{1}_{n_{s2}}^T \mathbf{x}^{s2}.
\end{align*}
\end{asmp}

\begin{thm}\label{thm_identifiability}
Under Assumption \ref{asmp_community}, for $\chi_1$ and $\chi_2$ defined in \eqref{eq_chi1_chi2}, $\chi_1 \not= \chi_2$ if and only if Assumption \ref{asmp_stubborn_agents} holds.
\end{thm}


\begin{algorithm}[!ht]
\caption{\newline Community Detection Algorithm for Gossip Dynamics}
\label{alg_1}
\textbf{Input: } States of regular agents $\{x^r(t), t\in \mathbb{N}\}$, initial vector of stubborn agents $x^s(0)$, and connection information between regular and stubborn agents (Assumption \ref{asmp_stubborn_agent_structure}).\\
\textbf{Output: } Community structure $\{\mathcal{C}(i), i \in \mathcal{V}\}$ and estimates of parameters $\hat{w}_s$ and $\hat{w}_d$.
\begin{algorithmic}[1]
\STATE{Randomize initial values for $\mathcal{C}(i,0)$, $\hat{w}_s(0)$, and $\hat{w}_d(0)$, and set $s^r(0) = x^r(0)$. }
\FOR{$t = 1, \dots$}
\STATE{Compute
\begin{align*}
    s^r(t) &= \frac{t}{t+1} s^r(t-1) + \frac{1}{t+1} x^r(t),\\
    \bar{s}^r(t) &= \frac{1}{|\mathcal{V}_r|}\mathbf{1}_{|\mathcal{V}_r|}^T s^r(t).
\end{align*}
}
\STATE{\textbf{Community detection part:}
\begin{align*}
    \hat{\mathcal{C}}(i, t) &= 2 - \mathbb{I}_{[s^r_i(t) > \bar{s}^r(t)]}, ~i \in \mathcal{V}_r,\\
    \hat{\mathcal{C}}(i, t) &= \hat{\mathcal{C}}(j_i, t), i \in \mathcal{V}_s,
\end{align*}
where $j_i$ is defined in Assumption \ref{asmp_stubborn_agent_structure}.
}
\STATE{\textbf{Parameter estimation part:}
\begin{align*}
    \tilde{w}_s(t) &= \hat{w}_s(t-1) -  \frac{1}{t} \text{sgn}(g(t)) \Big(g(t) \hat{w}_s(t-1) \\
    &\quad+ \frac{\beta_2(t)}{2\hat{n}_1(t)\hat{n}_2(t)}\Big),\\
    \hat{w}_s(t) &= \tilde{w}_s(t) \mathbb{I}_{[|\tilde{w}_s(t)| < 2]} + \frac12 \mathbb{I}_{[|\tilde{w}_s(t)| \ge 2]},\\
    \hat{w}_d(t) &= \frac{1 - (\hat{n}_1^2(t) + \hat{n}_2^2(t))\hat{w}_s(t)}{2\hat{n}_1(t)\hat{n}_2(t)},
\end{align*}
where 
\begin{align*}
    g(t) &= \beta_1(t) - \frac{\hat{n}_1^2(t) + \hat{n}_2^2(t)}{2\hat{n}_1(t)\hat{n}_2(t)} \beta_2(t),\\
    \beta_1(t) &= \frac{|\hat{\mathcal{V}}_{s1}(t)|}{|\hat{\mathcal{V}}_{r1}(t)|} \sum_{i\in \hat{\mathcal{V}}_{r1}(t)} s^r_i(t) - \sum_{i\in \hat{\mathcal{V}}_{s1}(t)} x_i(t),\\
    \beta_2(t) &= \frac{\hat{n}_2(t)}{|\hat{\mathcal{V}}_{r1}(t)|} \sum_{i\in \hat{\mathcal{V}}_{r1}(t)} s^r_i(t)\\
    &\quad  - \sum_{i\in \hat{\mathcal{V}}_{r2}(t)} s^r_i(t) - \sum_{i\in \hat{\mathcal{V}}_{s2}(t)} x_i(t),\\
    \hat{n}_k(t) &= \sum\nolimits_{i \in \mathcal{V}} \mathbb{I}_{[\hat{\mathcal{C}}(i,t) = k]},\\
    \hat{\mathcal{V}}_{rk}(t) &= \{i \in \mathcal{V}_r : \hat{\mathcal{C}}(i,t) = k\},\\
    \hat{\mathcal{V}}_{sk}(t) &= \{i \in \mathcal{V}_s : \hat{\mathcal{C}}(i,t) = k\}, ~k=1,2.
\end{align*}
}
\ENDFOR
\end{algorithmic}
\end{algorithm}

\begin{rmk}
The above theorem illustrates an intuitive but crucial fact that the average of states of stubborn agents in different communities must not be identical. Otherwise, their influence on regular agents in different communities would be the same, making it impossible to recover the community structure.
\end{rmk}

Since we have no information for the community structure of stubborn agents, we assume the following connections between them and regular ones. Intuitively, it means that we have some prior knowledge of stubborn agents' communities, which may be gathered from other source of data in practice.

\begin{asmp}\label{asmp_stubborn_agent_structure}
For every stubborn agent $i \in \mathcal{V}_s$, it is known that there exists a regular agent $j_i \in \mathcal{V}_r$ such that $\mathcal{C}(i) = \mathcal{C}(j_i)$.
\end{asmp}

Now we are ready to introduce the online community detection algorithm, as shown in Algorithm \ref{alg_1}, denoting the estimates of $\mathbf{x}^r$ in \eqref{eq_expectation_limit}, the community structure $\mathcal{C}(i)$, $w_s$, and $w_d$ at time $t$ by $s^r(t)$, $\hat{\mathcal{C}}(i,t)$, $\hat{w}_s(t)$, and $\hat{w}_d(t)$, $i \in \mathcal{V}$, respectively. The algorithm is based on ergodicity property \eqref{eq_ergodicity} of the system. The time average of each regular agent's trajectory is computed, and then used to cluster the agents into two groups. The idea is to split the agents by comparing the time averages of their states. Other clustering methods may be used to solve this problem, but it should be noted that the clustering result must have theoretical guarantees, to ensure consistent estimation of $w_s$ and $w_d$. Otherwise, incorrect knowledge of community structure could result in inconsistent estimation of parameters. 

Note from Theorem \ref{thm_stability} and \eqref{eq_update_compact_regular}, it follows that $\mathbf{x}^r$ satisfies the following equation, \begin{align*}
    \mathbf{x}^r = \bar{A} \mathbf{x}^r + \bar{B} \mathbf{x}^s,
\end{align*}
which implies that
\begin{align*}
    w_s(n_{s1}\chi_1 - \mathbf{1}_{n_{s1}}^T \mathbf{x}^{s1}) + w_d(n_{2}\chi_1 - n_{r2}\chi_2 - \mathbf{1}_{n_{s2}}^T \mathbf{x}^{s2}) = 0.
\end{align*}
Hence, $(n_{s1}\chi_1 - \mathbf{1}_{n_{s1}}^T \mathbf{x}^{s1})(n_{2}\chi_1 - n_{r2}\chi_2 - \mathbf{1}_{n_{s2}}^T \mathbf{x}^{s2}) < 0$. From the definition of $W$, $w_s$ and $w_d$ also have relation $w_s (n_1^2 + n_2^2) + 2 w_d n_1 n_2 = 1$, with $n_1^2 + n_2^2, n_1n_2 > 0$. 
Therefore, linear system
\begin{align*}
    \begin{cases}
    (n_{s1}\chi_1 - \mathbf{1}_{n_{s1}}^T \mathbf{x}^{s1})x + (n_{2}\chi_1 - n_{r2}\chi_2 - \mathbf{1}_{n_{s2}}^T \mathbf{x}^{s2})y = 0\\
    (n_1^2 + n_2^2) x + 2 n_1 n_2 y = 1
    \end{cases}
\end{align*}
has a unique solution $(w_s ~ w_d)$, given $n_1$ and $n_2$. There are multiple ways to solve the equation, and here we use a stochastic approximation algorithm, as presented in Line 5 of Algorithm \ref{alg_1}. 

We have the following result for Algorithm \ref{alg_1}, saying that the community detection task can be done in finite time, and the estimation of communication parameters is consistent.

\begin{thm}\label{thm_convergence}(Convergence of Algorithm \ref{alg_1})~\\
Suppose that Assumptions \ref{asmp_community}, \ref{asmp_stubborn_agents}, and \ref{asmp_stubborn_agent_structure} hold.\\
(i) The community detection part of Algorithm \ref{alg_1} converges in finite time.\\
(ii) The parameter estimation part of Algorithm \ref{alg_1} converges almost surely, i.e.,
\begin{align*}
    \mathbb{P}\left\{\lim_{t \to \infty}(\hat{w}_s(t), \hat{w}_d(t)) = (w_s, w_d)\right\} = 1,
\end{align*}
where $\hat{w}_s(t)$ and $\hat{w}_d(t)$ are the estimates of $w_s$ and $w_d$ at time $t$.
\end{thm}

\section{NUMERICAL SIMULATIONS}\label{sec_simulation}
In this section, we first illustrate convergence of the proposed algorithm by solving the problem for a gossip model with block structure in Assumption \ref{asmp_community}. Zachary's karate club network in the study of community detection is then used to test the performance of the algorithm.

\begin{figure}
    \centering
    \subfigure[\label{fig_community_detection}Finite-time convergence of the community detection part of Algorithm \ref{alg_1}.]{
    \includegraphics[width=0.6\linewidth]{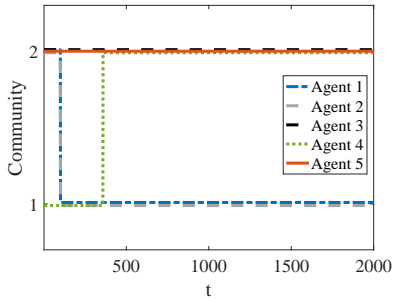}}
    \subfigure[\label{fig_parameter_estimation}Convergence of the parameter estimation part of Algorithm \ref{alg_1}, where the red lines are true values of $w_s$ and $w_d$.]{
    \includegraphics[width=0.6\linewidth]{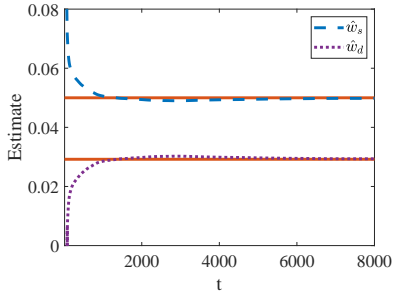}}
    \caption{\label{fig_1}Convergence of Algorithm \ref{alg_1}}
\end{figure}

To illustrate convergence of the proposed algorithm, consider a network consisting of five nodes, i.e., $\mathcal{V} = \{1,2,3,4,5\}$. Set $\mathcal{V}_{r1} = \{1\}$, $\mathcal{V}_{s1} = \{2\}$, $\mathcal{V}_{r2} = \{3, 4\}$, and $\mathcal{V}_{s2} = \{5\}$. In addition, let $w_s = 0.05$ and $w_d = 7/240$. The initial values of agents are drawn from independent standard Gaussian distribution. The performance of Algorithm \ref{alg_1} is shown in Fig. \ref{fig_1}. Finite-time convergence of the community detection algorithm can be observed in Fig. \ref{fig_community_detection}, and consistency of the parameter estimation part is demonstrated in Fig. \ref{fig_parameter_estimation}.

\begin{figure}
    \centering
    \subfigure[\label{fig_zachary_network}The community structure of Zachary's karate club network. Nodes drawn as red squares are associated with agent $1$, while nodes drawn as green triangles with agent $34$.]{
    \includegraphics[width=0.7\linewidth]{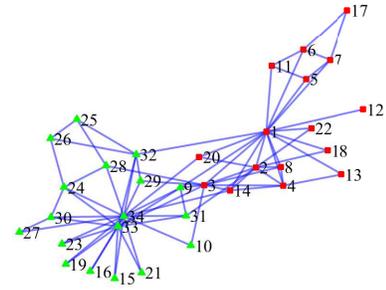}}
    \subfigure[\label{fig_accuracy}Accuracy of Algorithm \ref{alg_1} for gossip dynamics over Zachary's karate club network.]{
    \includegraphics[width=0.6\linewidth]{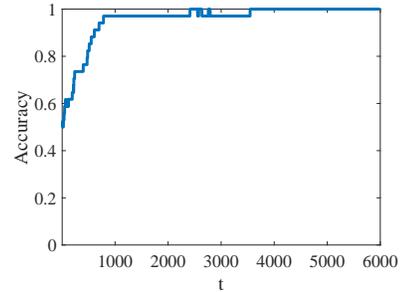}}
    \caption{\label{fig_2}Numerical experiment over Zachary's karate club network.}
\end{figure}

Zachary's karate club network \cite{zachary1977information}, containing $34$ members and presented in Fig. \ref{fig_zachary_network}, is used to demonstrate an application of the proposed algorithm. In \cite{zachary1977information}, a conflict between agents $1$ and $34$ resulted in fission of the club. It was shown that the network of friendships forecast the actual division of the group. In this numerical experiment, we suppose that an gossip opinion formation process takes place over the network, and agents $1$ and $34$ are the only stubborn agents in the network, holding different beliefs. At each time, one edge in Fig. \ref{fig_zachary_network} is selected with equal probability, and two agents corresponding to this edge communicate. The goal is to divide the group into partitions only based on the states of agents. Note that the network departures from our assumptions. The result is shown in Fig. \ref{fig_accuracy}, indicating that as time increases, our algorithm can finally recover the community structure of the group. The accuracy at time $t$ in Fig. \ref{fig_accuracy} is defined by $\frac1n \max_{\pi \in S_2} \{\sum_{i=1}^n \mathbb{I}_{[\mathcal{C}(i) = \pi(\hat{\mathcal{C}}(i,t))]}\}$, where $\pi : \{1, 2\} \to \{1, 2\}$ is a permutation function, $S_2$ is the group of permutations on $\{1, 2\}$, $\mathcal{C}(i)$ is agent $i$'s actual community, $\hat{\mathcal{C}}(i,t)$ is the estimate of agent $i$'s community at time $t$, and $n = 34$ in this numerical simulation.

\section{CONCLUSION}\label{sec_conclusion}
In this paper, we considered a community detection problem for gossip dynamics with stubborn agents. A community detection algorithm was proposed to recover community structure and also estimate communication probability parameters. It was proved that the community detection part converges in finite time, and the parameter estimation part converges almost surely. 

One of the ongoing works is to generalize the two-community assumption to multiple-community one, and to consider other types of assumptions similar to those in the study of stochastic block models \cite{abbe2017community}. Also, community detection problems of gossip systems with general graph structure and update rules will be considered in the future.

\bibliographystyle{IEEEtran}
\bibliography{IEEEabrv,bibliography.bib}

\begin{thebibliography}{10}
\providecommand{\url}[1]{#1}
\csname url@samestyle\endcsname
\providecommand{\newblock}{\relax}
\providecommand{\bibinfo}[2]{#2}
\providecommand{\BIBentrySTDinterwordspacing}{\spaceskip=0pt\relax}
\providecommand{\BIBentryALTinterwordstretchfactor}{4}
\providecommand{\BIBentryALTinterwordspacing}{\spaceskip=\fontdimen2\font plus
\BIBentryALTinterwordstretchfactor\fontdimen3\font minus
  \fontdimen4\font\relax}
\providecommand{\BIBforeignlanguage}[2]{{%
\expandafter\ifx\csname l@#1\endcsname\relax
\typeout{** WARNING: IEEEtran.bst: No hyphenation pattern has been}%
\typeout{** loaded for the language `#1'. Using the pattern for}%
\typeout{** the default language instead.}%
\else
\language=\csname l@#1\endcsname
\fi
#2}}
\providecommand{\BIBdecl}{\relax}
\BIBdecl

\bibitem{fortunato2016community}
S.~Fortunato and D.~Hric, ``Community detection in networks: A user guide,''
  \emph{Physics Reports}, vol. 659, pp. 1--44, 2016.

\bibitem{schaub2017many}
M.~T. Schaub, J.-C. Delvenne, M.~Rosvall, and R.~Lambiotte, ``The many facets
  of community detection in complex networks,'' \emph{Applied Network Science},
  vol.~2, no.~1, pp. 1--13, 2017.

\bibitem{fortunato2010community}
S.~Fortunato, ``Community detection in graphs,'' \emph{Physics Reports}, vol.
  486, no. 3-5, pp. 75--174, 2010.

\bibitem{von2007tutorial}
U.~Von~Luxburg, ``A tutorial on spectral clustering,'' \emph{Statistics and
  Computing}, vol.~17, no.~4, pp. 395--416, 2007.

\bibitem{newman2004finding}
M.~E. Newman and M.~Girvan, ``Finding and evaluating community structure in
  networks,'' \emph{Physical Review E}, vol.~69, no.~2, p. 026113, 2004.

\bibitem{abbe2017community}
E.~Abbe, ``Community detection and stochastic block models: recent
  developments,'' \emph{The Journal of Machine Learning Research}, vol.~18,
  no.~1, pp. 6446--6531, 2017.

\bibitem{prokhorenkova2019learning}
L.~Prokhorenkova, A.~Tikhonov, and N.~Litvak, ``Learning clusters through
  information diffusion,'' in \emph{The World Wide Web Conference}, 2019, pp.
  3151--3157.

\bibitem{peixoto2019network}
T.~P. Peixoto, ``Network reconstruction and community detection from
  dynamics,'' \emph{Physical Review Letters}, vol. 123, no.~12, p. 128301,
  2019.

\bibitem{wai2019blind}
H.-T. Wai, S.~Segarra, A.~E. Ozdaglar, A.~Scaglione, and A.~Jadbabaie, ``Blind
  community detection from low-rank excitations of a graph filter,'' \emph{IEEE
  Transactions on Signal Processing}, 2019.

\bibitem{schaub2019blind}
M.~T. Schaub, S.~Segarra, and J.~N. Tsitsiklis, ``Blind identification of
  stochastic block models from dynamical observations,'' \emph{arXiv preprint
  arXiv:1905.09107}, 2019.

\bibitem{roddenberry2020exact}
T.~M. Roddenberry, M.~T. Schaub, H.-T. Wai, and S.~Segarra, ``Exact blind
  community detection from signals on multiple graphs,'' \emph{arXiv preprint
  arXiv:2001.10944}, 2020.

\bibitem{ramezani2018community}
M.~Ramezani, A.~Khodadadi, and H.~R. Rabiee, ``Community detection using
  diffusion information,'' \emph{ACM Transactions on Knowledge Discovery from
  Data (TKDD)}, vol.~12, no.~2, pp. 1--22, 2018.

\bibitem{fagnani2008randomized}
F.~Fagnani and S.~Zampieri, ``Randomized consensus algorithms over large scale
  networks,'' \emph{IEEE Journal on Selected Areas in Communications}, vol.~26,
  no.~4, pp. 634--649, 2008.

\bibitem{acemouglu2013opinion}
D.~Acemo{\u{g}}lu, G.~Como, F.~Fagnani, and A.~Ozdaglar, ``Opinion fluctuations
  and disagreement in social networks,'' \emph{Mathematics of Operations
  Research}, vol.~38, no.~1, pp. 1--27, 2013.

\bibitem{zachary1977information}
W.~W. Zachary, ``An information flow model for conflict and fission in small
  groups,'' \emph{Journal of Anthropological Research}, vol.~33, no.~4, pp.
  452--473, 1977.

\end{thebibliography}






\end{document}